\documentclass[conference]{IEEEtran}
\IEEEoverridecommandlockouts
\usepackage{cite}
\usepackage{amsmath,amssymb,amsfonts}

\usepackage{graphicx}
\usepackage{textcomp}
\usepackage{xcolor}
\usepackage{braket}
\usepackage{hyperref}
 \usepackage{booktabs} 
 \usepackage{algorithm}
\usepackage{algpseudocode}
\usepackage[caption=false,font=footnotesize]{subfig} 
\usepackage{amsmath,amssymb}

\hypersetup{
    colorlinks=true, % Use colored links instead of boxes
    linkcolor=black,  % Color of internal links
    citecolor=black, % Color of citation links
    filecolor=black, % Color of file links
    urlcolor=black    % Color of external links
}

\def\BibTeX{{\rm B\kern-.05em{\sc i\kern-.025em b}\kern-.08em
    T\kern-.1667em\lower.7ex\hbox{E}\kern-.125emX}}
\begin{document}

\title{Distributed Quantum Neural Networks on Distributed Photonic Quantum Computing
}

\author{
\IEEEauthorblockN{
    Kuan-Cheng Chen\IEEEauthorrefmark{2}\IEEEauthorrefmark{3}\IEEEauthorrefmark{1},
    Chen-Yu Liu\IEEEauthorrefmark{4}\IEEEauthorrefmark{5}, 
    Yu Shang\IEEEauthorrefmark{6}\IEEEauthorrefmark{3},
    Felix Burt\IEEEauthorrefmark{2}\IEEEauthorrefmark{3},
    Kin K. Leung \IEEEauthorrefmark{2}
}

\IEEEauthorblockA{\IEEEauthorrefmark{2}Department of Electrical and Electronic Engineering, Imperial College London, London, UK}
\IEEEauthorblockA{\IEEEauthorrefmark{3}Centre for Quantum Engineering, Science and Technology (QuEST), Imperial College London, London, UK}
\IEEEauthorblockA{\IEEEauthorrefmark{4}Graduate Institute of Applied Physics, National Taiwan University, Taipei, Taiwan}
\IEEEauthorblockA{\IEEEauthorrefmark{6}Blackett Laboratory, Department of Physics, Imperial College London, London, UK}
\IEEEauthorblockA{Email:\IEEEauthorrefmark{1} kuan-cheng.chen17@imperial.ac.uk, \IEEEauthorrefmark{5} d10245003@g.ntu.edu.tw}
}

\maketitle

\begin{abstract}
We introduce a distributed quantum-classical framework that synergizes photonic quantum neural networks (QNNs) with matrix product state (MPS) mapping to achieve parameter-efficient training of classical neural networks. By leveraging universal linear-optical decompositions of $M$-mode interferometers and photon-counting measurement statistics, our architecture generates neural parameters through a hybrid quantum-classical workflow: photonic QNNs with $M(M+1)/2$ trainable parameters produce high-dimensional probability distributions that are mapped to classical network weights via an MPS model with bond dimension $\chi$. Empirical validation on MNIST classification demonstrates that photonic QT achieves an accuracy of \(95.50\% \pm 0.84\%\) using 3,292 parameters (\(\chi = 10\)), compared to \(96.89\% \pm 0.31\%\) for classical baselines with 6,690 parameters. Moreover, a ten-fold compression ratio is achieved at \(\chi = 4\), with a relative accuracy loss of less than \(3\%\). The framework outperforms classical compression techniques (weight sharing/pruning) by 6-12\% absolute accuracy while eliminating quantum hardware requirements during inference through classical deployment of compressed parameters. Simulations incorporating realistic photonic noise demonstrate the framework’s robustness to near-term hardware imperfections. Ablation studies confirm quantum necessity - replacing photonic QNNs with random inputs collapses accuracy to chance level ($10.0\% \pm 0.5\%$). Photonic quantum computing room-temperature operation, inherent scalability through spatial mode multiplexing, and HPC-integrated architecture establish a practical pathway for distributed quantum machine learning, combining the expressivity of photonic Hilbert spaces with the deployability of classical neural networks.
\end{abstract}


\section{Introduction}
Quantum-centric supercomputing represents a transformative paradigm integrating classical high-performance computing (HPC) architectures with distributed quantum resources to overcome fundamental limitations of isolated quantum systems \cite{bravyi2022future,gambetta2022quantum}. While classical architectures excel in high-precision numerical computation and massive data throughput, quantum processors enable navigation of exponentially high-dimensional Hilbert spaces ($\mathcal{H} \sim \mathbb{C}^{2^n}$ for $n$ qubits) and optimization of non-convex cost landscapes through quantum parallelism \cite{biamonte2017quantum,schuld2019quantum}. This hybrid architecture proves particularly effective for quantum machine learning (QML), enabling: quantum-enhanced feature mapping via parameterized quantum circuits\cite{benedetti2019parameterized,dunjko2016quantum,chen2024validating,chen2024quantumsvm}, classical processing of measurement outcomes through neural networks\cite{huang2021power,jerbi2024shadows}, and co-optimization of hybrid quantum-classical objective functions \cite{ciliberto2018quantum,mcclean2016theory,caro2022generalization,cerezo2022challenges}. The paradigm addresses current quantum hardware constraints by leveraging classical HPC for error mitigation, resource allocation across distributed quantum nodes, and real-time calibration of noisy intermediate-scale quantum (NISQ) processors \cite{cerezo2021variational,preskill2018quantum}. Therefore, a noise-robustness framework is a key contribution to enabling practical near-term applications.

Compared to other quantum computing techniques, photonic quantum computing emerges as a cornerstone for distributed quantum–classical workflows due to its inherent scalability, room-temperature operation, and compatibility with existing optical networks \cite{wang2020integrated,alexander2024manufacturable,larsen2021deterministic,yu2024shedding}.
Qubits encoded in photonic degrees of freedom (e.g., polarization, time bins, or spatial modes) resist decoherence from thermal noise, enable low-loss transmission over fiber-optic channels, and support wavelength-division multiplexing for parallel processing\cite{slussarenko2019photonic}.
These traits position photonic systems as natural candidates for distributed quantum computing (DQC), where spatially separated quantum processors collaborate via classical/quantum communication to solve problems intractable on monolithic quantum devices\cite{main2025distributed,caleffi2024distributed}.

\begin{figure}[!t]
\centering
\includegraphics[width=1.05\linewidth]{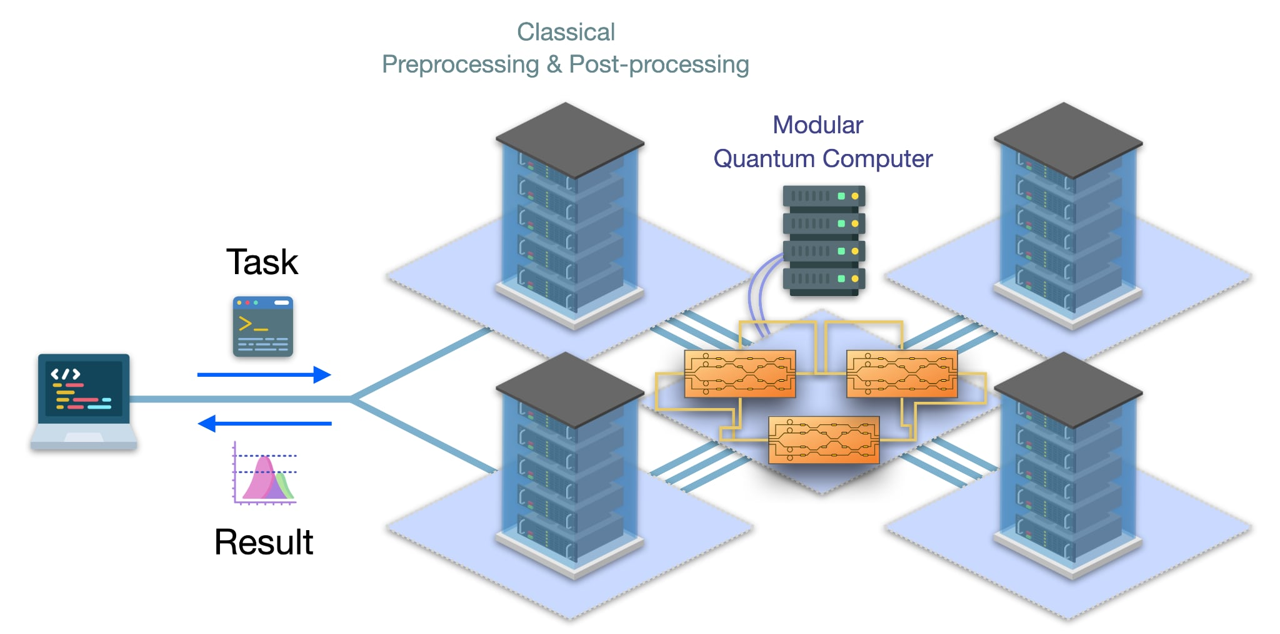}
\caption{
Schematic of quantum-centric supercomputing based on a distributed photonic quantum-computing architecture.
}
 \label{fig:demo}
\end{figure}

In conventional QML frameworks, quantum neural networks (QNNs) encode classical data into parameterized quantum circuits, with gradients computed on classical HPC systems\cite{chen2020variational,yu2024QONN}.
However, centralized quantum processing introduces critical bottlenecks: (i)~high-dimensional data embedding necessitates deep circuits prone to decoherence\cite{huang2022quantum,chia2023need}; (ii)~synchronous access to quantum hardware imposes latency incompatible with real-time applications\cite{huang2021power}; and (iii)~scaling to large models exacerbates the \textit{memory wall} between quantum and classical subsystems\cite{perez2020data}.
QML approaches based on distributed architectures mitigate these challenges by partitioning or offloading large computational tasks or unitary operations into smaller subproblems, leveraging multiplexing techniques to parallelize parameterized quantum operations across multiple smaller, yet higher-fidelity, quantum processors \cite{sheng2017distributed,chen2021federated,hwang2024distributed,pira2023invitation,ma2025robust}.

\begin{figure*}[ht]
\centering
\includegraphics[width=\linewidth]{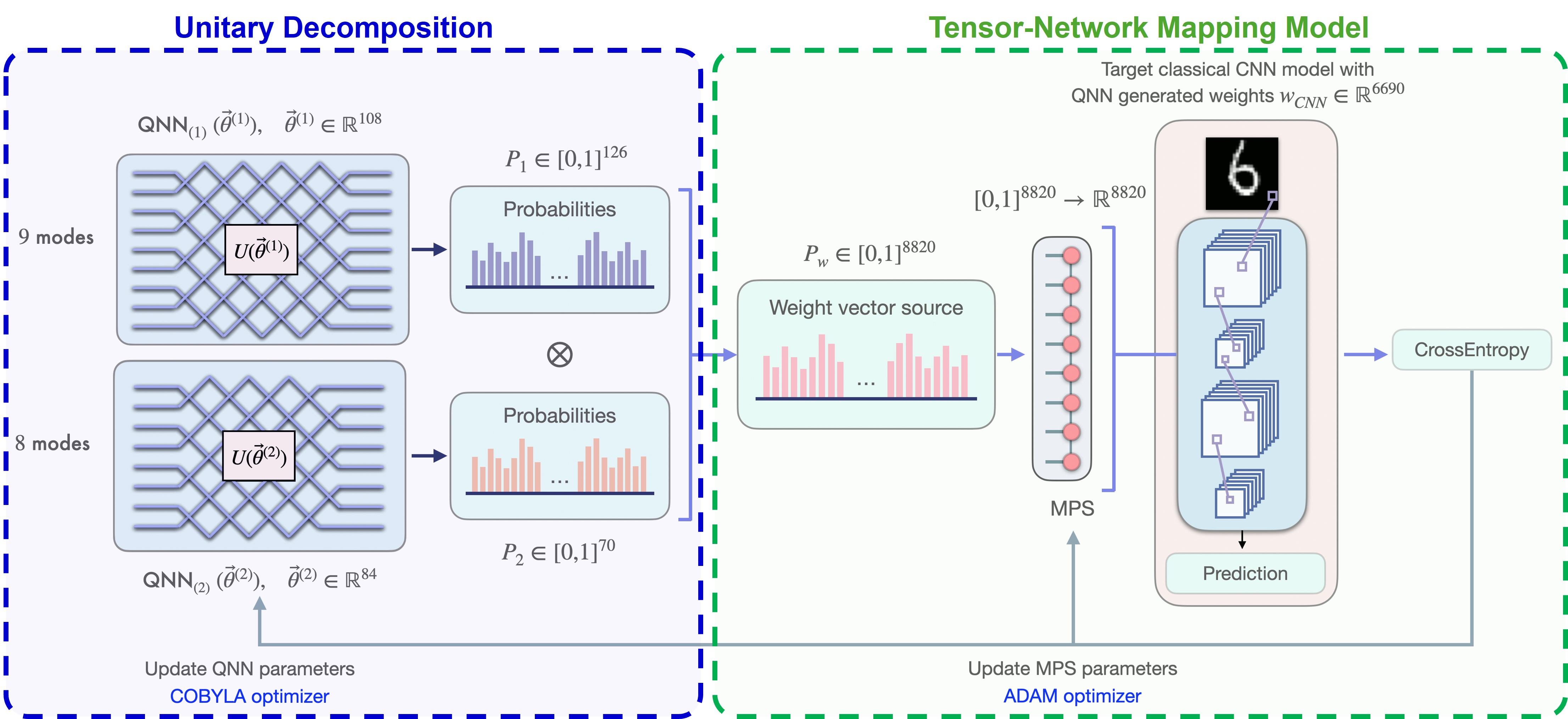}
\caption{Overview of the Distributed Photonic Quantum Neural Network (DQNN) framework with unitary decomposition and MPS-based mapping.
    The system comprises two PQCs, $\text{QNN}_{(1)}$ and $\text{QNN}_{(2)}$, consisting of 9 and 8 optical modes, respectively. Each PQC encodes a trainable unitary $U(\vec{\theta}^{(i)}) \in \mathrm{U}(M_i)$ implemented via a universal decomposition into beam-splitters and phase shifters. Upon injecting $N_i$ photons into $M_i$-mode interferometers, the resulting measurement statistics $P_1 \in [0,1]^{126}$ and $P_2 \in [0,1]^{70}$ span a high-dimensional simplex. These probability vectors are combined via a tensor product $P_w = P_1 \otimes P_2 \in [0,1]^{8820}$ to form the source of weight candidates for a classical convolutional neural network (CNN). The mapping from $[0,1]^{8820}$ to $\mathbb{R}^{8820}$ is performed using a Matrix Product State (MPS) model, $G_b$, which is trained to project $P_w$ into a lower-dimensional subspace containing the target CNN weights $w_{\text{CNN}} \in \mathbb{R}^{6690}$. The CNN is trained to classify inputs using the QNN-generated weights and is evaluated with a cross-entropy loss. Optimization is conducted using the COBYLA algorithm for quantum circuit parameters $\vec{\theta}^{(i)}$, and the ADAM optimizer for classical MPS parameters $b$.}
\label{fig:scheme}
\end{figure*}

To advance the integration of quantum computing with HPC, Liu et al.\cite{liu2024quantum} introduce the \emph{Quantum-Train} (QT) framework. This framework exemplifies a distributed paradigm by employing QNNs as quantum hyper-networks that generate weights for classical neural networks (NNs). Notably, this approach inverts the traditional roles in quantum machine learning (QML): quantum resources are utilized to optimize model parameters during training, while inference is performed entirely on classical hardware. However, previous implementations of QT on gate-based quantum platforms—such as superconducting and trapped-ion systems—have faced scalability challenges, primarily due to cryogenic infrastructure requirements and limited qubit connectivity\cite{liu2024qtrl,lin2024quantum,liu2025federated,chen2024quantum,liu2024programming,chen2025toward}.

Toward extending the QT framework to photonic architectures (namely, photonic-QT), we exploit several native advantages. First, distributed scalability is enabled by encoding optical qubits in spatially separated modes, allowing modular deployment of QNNs across multiple photonic processors\cite{aghaee2025scaling}. Second, HPC integration is facilitated as classical HPC clusters coordinate outputs from photonic QNNs via tensor-network contractions, effectively bridging quantum and classical parameter spaces. Third, noise resilience is achieved because photonic qubits are intrinsically resistant to thermal decoherence, while measurement-based feedback mechanisms correct photon-loss errors \cite{langford2011efficient}.

In this research work, we introduce a distributed learning QT framework that synergizes photonic QNNs with classical HPC orchestration.
For example, two photonic processors—each hosting \(N\) qubits in \(M\) multiplexed optical modes—generate photon-count probabilities \(p_\theta(y)\) mapped via tensor networks to convolutional neural network (CNN) weights. Classical HPC resources optimize the photonic parameters \(\theta\) through backpropagation, while a distributed scheduler allocates quantum tasks across photonic nodes to minimize communication overhead (Fig.,\ref{fig:scheme}).
This architecture achieves three advances:

\begin{itemize}
\item \textbf{Distributed quantum resource scaling.} By partitioning QNN workloads across photonic processors, photonic QT reduces per-node qubit requirements to \(O(NM)\).
\item \textbf{Decentralized photonic–HPC integration.} Quantum parameter updates occur asynchronously across photonic nodes, with classical HPC aggregating gradients via consensus protocols,\cite{lin2024quantum}. This eliminates latency bottlenecks associated with centralized quantum control.
\item \textbf{Task-agnostic photonic compatibility.} The framework’s separation of quantum training and classical inference extends to recurrent networks, federated learning, and reinforcement tasks,\cite{liu2024federated,liu2024programming}, demonstrating compatibility with diverse QML workloads.
\end{itemize}

By unifying photonic quantum computing’s scalability with distributed HPC coordination, our strategy establishes a blueprint for fault-tolerant\cite{takeda2019toward}, large-scale quantum–classical architectures.
Our experimental validation on MNIST classification benchmarks confirms that photonic QNNs can match gate-based QT performance while leveraging optical parallelism to accelerate training cycles.
This work underscores photonic technologies as a viable substrate for DQC.

\section{Related Work}
\label{sec:related}

\textbf{Parameter-efficient training in the classical domain.}  
Understanding the benefit of the proposed photonic‐QT scheme requires first surveying classical approaches that shrink neural-network models during optimization, rather than via post-hoc compression.  A rich literature explores such strategies, including unstructured or structured pruning\,\cite{blalock2020state}, parameter tying and weight sharing\,\cite{nowlan2018simplifying}, low-rank matrix and tensor factorisations\,\cite{kolda2009tensor}, knowledge distillation\,\cite{gou2021knowledge}, and low-precision quantisation\,\cite{lin2016fixed}.  Among these, pruning and weight sharing most closely match the objective of our photonic-QT framework—namely, reducing the number of trainable parameters without sacrificing expressive power.  In the experimental section we therefore benchmark photonic-QT against representative algorithms from these two families, reporting both quantitative compression metrics and qualitative generalisation behaviour.

\textbf{Quantum resources for training classical neural networks.}  
A complementary research strand exploits quantum processors solely within the training loop, thereby avoiding the overhead of quantum data encoding and dispensing with quantum hardware at inference time.  Early proposals range from quantum-walk search for weight optimisation\,\cite{de2021classical} to quantum hyper-networks that generate binary weights for compact classical models\,\cite{carrasquilla2023quantum}.  While conceptually appealing, the latter approach confines the trained networks to binary precision, narrowing its applicability.  By contrast, the Quantum-Train paradigm—and the photonic realisation advanced in this work—furnishes a more general mechanism: an $n$-qubit, polynomial-depth quantum neural network acts as a hyper-network that outputs full-precision weights for arbitrarily large classical NNs.  The result is a parameter-efficient, distribution-agnostic training protocol that (i) confines quantum computation to the offline training phase, (ii) retains a fully classical model for deployment, and (iii) is naturally amenable to room-temperature, multiplexed photonic hardware.  These properties render photonic-QT a scalable and practically viable candidate for next-generation distributed quantum–classical machine-learning pipelines.

\section{Unitary Decomposition-Based Parameter Compression}
\label{sec:QPA}
In this section, we propose a novel strategy that leverages Photonic Quantum Circuit (PQC)-based QNNs combined with a classical mapping model as an efficient parameter generator (see Algorithm \ref{alg:PQT}). The central insight underpinning our approach is that PQCs inherently provide high-dimensional Hilbert spaces, enabling compact representation of neural network parameters. Specifically, a limited number of QNN parameters can effectively control exponentially larger sets of measurement probabilities, offering a significant dimensional compression relative to classical neural networks.

We introduce a parameter generation scheme distinctly different from traditional QML. Given a target classical neural network with parameters $w_{\text{CNN}} = (w_1, w_2, \dots, w_m)$, we employ two photonic QNNs, denoted as $\text{QNN}{(1)}(\vec{\theta}^{(1)})$ and $\text{QNN}{(2)}(\vec{\theta}^{(2)})$, each configured with respective photon modes and qubit counts $(M_1, N_1)$ and $(M_2, N_2)$. The number of distinct measurement outcomes, determined by unitary decompositions of multi-mode interferometers, is given in combination as $C(M_i, N_i) = \frac{M_i!}{N_i!(M_i - N_i)!}$, for $i = 1,2$. These configurations are selected to satisfy the inequality:

\begin{equation}
\label{eq:nm_condition}
    C(M_1, N_1) \times C(M_2, N_2) \ge m.
\end{equation}

which ensures a substantial reduction in the number of parameters to be optimized directly within the quantum systems compared to the target neural network.

Measurement probabilities from each QNN, denoted as vectors $P_1 \in [0,1]^{C(M_1, N_1)}$ and $P_2 \in [0,1]^{C(M_2, N_2)}$, are tensorially combined to form a high-dimensional parameter vector:

\begin{equation}
    P_w = P_1 \otimes P_2 \in [0,1]^{C(M_1, N_1) \times C(M_2, N_2)}.
\end{equation}

Remarkably, by controlling only $\frac{3M_1(M_1-1)}{2} + \frac{3M_2(M_2-1)}{2}$ parameters—originating from unitary decompositions of linear-optical interferometers in each QNN—we can generate sufficient parameters to populate $w_{\text{CNN}}$. As $P_w \in [0,1]$, while $w_{\text{CNN}}$ typically spans the real domain $\mathbb{R}$, an additional classical mapping model $G_{b}$, parameterized by $b$, performs the necessary mapping $[0,1] \rightarrow \mathbb{R}$. Following methodologies outlined in \cite{liu2025quantumTN}, we adopt a Matrix Product State (MPS)-based mapping model, yielding:

\begin{equation}
    w_{\text{CNN}} \subseteq G_{b}(P_w) \in \mathbb{R}^{C(M_1, N_1) \times C(M_2, N_2)},
\end{equation}

where inclusion indicates that redundant generated parameters are discarded after fulfilling the exact parameter count $m$. By adjusting parameters $(\vec{\theta}^{(1)}, \vec{\theta}^{(2)}, b)$, we optimize the loss function $\mathcal{L}$ evaluated by the target neural network.

\subsection{Theoretical Feasibility via Universal Linear–Optical Decomposition}

Any unitary $U^{(i)}\!\in\!\mathrm U(M_i)$ underlying $\text{QNN}^{(i)}$ admits a Clements (or Reck) factorisation into
$L_i=M_i(M_i-1)/2$ Mach–Zehnder interferometers $B_{p_lq_l}(\theta_l)$ interleaved with $M_i$ phase shifters
$\Phi_k(\varphi_k)$:\footnote{The Clements decomposition is depth-optimal for planar layouts\cite{clements2016optimal}.}
\begin{equation}
U^{(i)}=\Bigl[\prod_{l=1}^{L_i} B_{p_lq_l}(\theta_l)\Bigr]
        \Bigl[\prod_{k=1}^{M_i}\Phi_k(\varphi_k)\Bigr].
\label{eq:unitary-decomp}
\end{equation}
Thus the number of continuous, trainable parameters for one $M_i$-mode interferometer is  
\begin{equation}
\label{eq:trainable-count}
N_{\mathrm{train}}^{(i)} \;=\; L_i + M_i
                             \;=\; \frac{M_i(M_i+1)}{2}.
\end{equation}
Equation~\eqref{eq:unitary-decomp} therefore proves that the total number of continuous parameters is exactly
$N_{\mathrm{train}}^{(i)}$, matching \eqref{eq:trainable-count} and confirming that our ansatz exploits \emph{all}
degrees of freedom afforded by universal linear optics. Consequently, the probability vector $P_i$ is an injective function of $\vec\theta^{(i)}$, ensuring that every adjustment of the optical phases produces a distinct point in the probability simplex.  Combining two such unitaries across spatially separated photonic processors therefore spans a parameter space asymptotically larger than that of the target classical model whenever Condition Eq.~\eqref{eq:nm_condition} holds, establishing theoretical feasibility of the DQNN framework.

\subsection{Gradient Computation for Quantum-Compressed Neural Parameters.}

The gradients of the loss function with respect to the quantum parameters $(\vec{\theta}^{(i)}, b)$ require special consideration due to the quantum-classical hybrid structure. They are computed through the Jacobian of the classical parameters relative to quantum parameters:

\begin{equation}
\label{eq:grad}
\nabla_{\vec{\theta}^{(i)}, b} \mathcal{L} = \left(\frac{\partial w_{\text{CNN}}}{\partial (\vec{\theta}^{(i)}, b)}\right)^T \cdot \nabla_{w_{\text{CNN}}} \mathcal{L}.
\end{equation}

Here, the Jacobian captures sensitivities induced by quantum measurement probabilities, reflecting dependencies established through unitary decompositions of linear optics elements.

\subsection{Parameter Update of Photonic Quantum Circuit Compressed Parameters.} 

The learning rate $\eta$ is a critical factor, particularly given the complex dynamics introduced by the quantum-classical interface. The update rule for the quantum parameters is defined as:
\begin{equation}
\label{eq:update}
\vec{\theta}^{(i)}_{t+1}, b_{t+1} = \vec{\theta}^{(i)}_t, b_t + \eta \nabla_{\vec{\theta}^{(i)}, b} \mathcal{L}. %(\va_t).
\end{equation}
This update ensures that the quantum parameters are optimized to improve the performance of the target NN. The equation provides a high-level representation of the gradient update in an exact quantum state simulation. However, in practical applications using real quantum hardware or specific backend providers, the gradient calculation must incorporate the parameter shift rule and its variants \cite{mitarai2018quantum, schuld2019evaluating}.

\begin{algorithm}[H]
\caption{Photonic Quantum-Train Forward Pass}\label{alg:PQT}
\begin{algorithmic}[1]
\Require Input tensor $\mathbf{x} \in \mathbb{R}^{H\times W\times C}$ (height $\times$ width $\times$ channels),
         PQC parameters $\theta \in \mathbb{R}^d$,
         Photonic modes $\{M_k\}_{k=1}^K$,
         Measurement budget $N_{\text{samp}}$,
         CNN template $\mathcal{S} = \{(n_j,\mathbf{s}_j)\}$ (param counts $\times$ shapes)
\Ensure Class logits $\mathbf{y} \in \mathbb{R}^K$

\Statex \textsc{\textcolor{blue}{Photonic Parameter Generation}}
\State $\{\theta^{(k)}\}_{k=1}^K \gets \operatorname{split}(\theta)$ \Comment{Clements-decomposed MZI meshes}
\For{$k \in \{1,...,K\}$} \Comment{Parallel QNN execution}
    \State $\mathbf{p}^{(k)} \gets \operatorname{norm}(\operatorname{QNN}_k(\theta^{(k)}, N_{\text{samp}}))$ 
           \Comment{$C(M_k,N_k)$-dim probabilities per Eq.~(1)}
\EndFor
\State $\mathbf{P}_w \gets \operatorname{vec}(\bigotimes_{k=1}^K \mathbf{p}^{(k)})_{1:m}$ 
       \Comment{Truncated to $m$ params via Eq.~\eqref{eq:nm_condition}}

\Statex \textsc{\textcolor{blue}{Quantum-Classical Mapping}}
\State $\mathbf{v} \gets \operatorname{MPS}(\mathbf{P}_w; \chi)$ \Comment{$G_b$ with $\chi$ from Table~\ref{tab:setup}}
\State $\mathbf{v} \gets \mathbf{v} - \mu_{\mathbf{v}}$ \Comment{Centering}

\Statex \textsc{Parameter Allocation}
\State $\mathcal{W} \gets \{\}$; $c \gets 0$
\For{$(n_j, \mathbf{s}_j) \in \mathcal{S}$}
    \State $\mathcal{W}[j] \gets \operatorname{reshape}(\mathbf{v}[c{:}c{+}n_j], \mathbf{s}_j)$ 
           \Comment{Param slicing with $c \leq m$}
    \State $c \gets c + n_j$
\EndFor

\Statex \textsc{\textcolor{blue}{Classical Inference}}
\State $\mathbf{y} \gets \operatorname{Sequential}\big($
\State $\quad \operatorname{Conv2D}(\mathcal{W}[1]),\operatorname{MaxPool},$
\State $\quad \operatorname{Conv2D}(\mathcal{W}[2]),\operatorname{AvgPool},$
\State $\quad \operatorname{Flatten},\operatorname{Linear}(\mathcal{W}[3]),$
\State $\quad \operatorname{ReLU},\operatorname{Linear}(\mathcal{W}[4])\big)(\mathbf{x})$
\State \Return $\mathbf{y}$
\end{algorithmic}
\end{algorithm}

In our implementation, the mapping model parameters are updated using the ADAM optimizer, while the QNN parameters are updated using the COBYLA optimizer. An overview of the photonic QT framework is provided in Fig.~\ref{fig:scheme}, with detailed parameter settings described in the subsequent section.

\begin{table*}[htbp]
    \centering
    \caption{Configuration of the mapping model $G_{b}$.}
    \vspace{10pt} 
    \small
    \begin{tabular}{llcc}
        \toprule
          \textbf{Hyperparameter} & \textbf{Meaning} & \textbf{Value} \\
        \midrule
         Input size & Input of the mapping model $(| \phi_i \rangle, |\langle \phi_i | \psi (\vec{\theta}^{(i)}) \rangle|^2)$ \cite{liu2025quantumTN} & $\lceil \log_2 m \rceil +1$   \\
         Bond dimension & Main structure parameter of the MPS mapping model & $1 \sim 10$  \\
        \bottomrule
    \end{tabular}
    \label{tab:setup}
\end{table*}
 
\section{Empirical Experiments}
\label{sec:exp}

\subsection{Distributed Photonic Architecture}
In this study, we adopt the Clements decomposition as the foundation of our photonic QNN design, leveraging a multi-mode interferometer composed of phase shifters (PSs) and balanced beam splitters (BSs) arranged in a rectangular mesh \cite{clements2016optimal}. This systematic approach allows us to implement an arbitrary \( m \times m \) unitary transformation \( U \), an essential requirement for training flexible and reconfigurable photonic QNNs. Formally, the decomposition factorizes \( U \) into a product of two-mode operations:
\begin{equation}
    U = \prod_{l=1}^{L} B_{\theta_l, \phi_l},
\end{equation}
where each \( B_{\theta_l, \phi_l} \) denotes a two-mode beam splitter parameterized by the real variables \(\theta_l\) (transmittance or reflectance) and \(\phi_l\) (phase shift). An additional on-mode phase shifter follows each beam splitter to fine-tune the local optical phase. The total number of layers \( L \) scales on the order of \( m(m-1)/2 \), guaranteeing that \emph{any} desired transformation is realizable for a given device size.

\begin{figure}[!t]
\centering
\includegraphics[width=\linewidth]{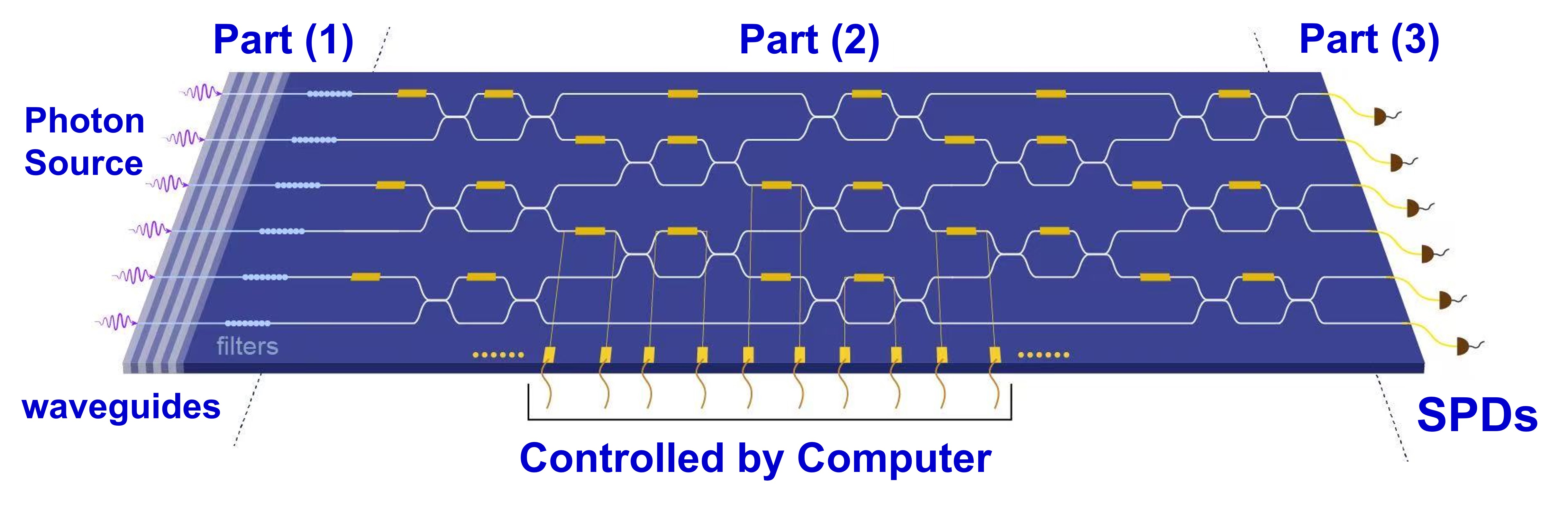}
\caption{
Schematic of a reconfigurable integrated photonic circuits. The device is structured into three main parts: (1) photon generation and initialization using on-chip photon sources and spectral filters; (2) quantum state evolution through a programmable interferometric network of tunable beam splitters and phase shifters, dynamically controlled by a software-controlled power supply; and (3) Photon detection using superconducting nanowire single-photon detectors (SNSPDs) or photon number resolution detectors (PNRDs).
}
 \label{fig:photonic-qc}
\end{figure}

Fig.~\ref{fig:scheme} (conceptual illustration) outlines our experimental workflow. We begin by initializing \( m \) photonic input modes in the quantum states required for the QNN algorithm (e.g., single photons per mode). These inputs then traverse a series of alternating BS and PS elements arranged in a checkerboard pattern. In each layer, the set of parameters \(\{\theta_l, \phi_l\}\) controls both the beam-splitter transmission coefficients and the additional local phases. Concretely, if \(\hat{a}_i^\dagger\) and \(\hat{a}_j^\dagger\) are the creation operators for two modes in a beam-splitter interaction at layer \( l \), this operation is given by:
\begin{equation}
    \begin{pmatrix}
        \hat{a}_i^\dagger \\
        \hat{a}_j^\dagger
    \end{pmatrix}
    \longrightarrow
    \begin{pmatrix}
        \cos \theta_l & - e^{-i\phi_l} \sin \theta_l \\
        e^{i\phi_l} \sin \theta_l & \cos \theta_l
    \end{pmatrix}
    \begin{pmatrix}
        \hat{a}_i^\dagger \\
        \hat{a}_j^\dagger
    \end{pmatrix}.
\end{equation}

Subsequent phase shifts are applied to each output mode via thermo-optic or electro-optic modulators. By iterating this sequence across all layers in the rectangular mesh, we achieve an \textit{in situ} realization of the global unitary \( U \), enabling the arbitrary multi-mode transformations \cite{yu2023gbs} critical to QNN training.

Our experimental setup builds upon the Perceval library \cite{heurtel2023perceval}, which programmatically allocates and updates the parameters \(\{\theta_l, \phi_l\}\) for each beam-splitter and phase-shifter element. Users may specify these parameters directly or allow randomized initialization for a gradient-based training protocol. This flexibility paves the way for closed-loop optimization of photonic QNN architectures under real-time feedback, significantly enhancing the practicality of quantum-enhanced machine learning and other quantum computing applications.

In addition to its reconfigurability, this photonic platform inherently operates at room temperature, exhibits comparatively lower noise, and scales readily via integrated photonic circuits. As a result, the Clements decomposition offers a robust and hardware-efficient framework to implement the distributed quantum neural network ansatz, setting the stage for the experiments and performance evaluations described in subsequent sections.

\subsection{Classical Baselines for Benchmarking}
\label{sec:baselines}

\begin{table*}[!t]
    \centering
    \caption{Result of the original classical CNN.}
    \vspace{10pt} 
    \small
    \begin{tabular}{cccc}
        \toprule
          \textbf{\# of training} & \textbf{Training} & \textbf{Testing} & \textbf{Generalization} \\
          \textbf{parameters} & \textbf{accuracy (\%)} & \textbf{accuracy (\%)} & \textbf{error} \\
        \midrule
          6690 & 99.983 $\pm$ 0.02 & 96.890 $\pm$ 0.31 & 0.1690 $\pm$ 0.005\\
        \bottomrule
    \end{tabular}
    \label{tab:original_cnn}
\end{table*}

\begin{figure*}[!t]
\centering
\includegraphics[width=\linewidth]{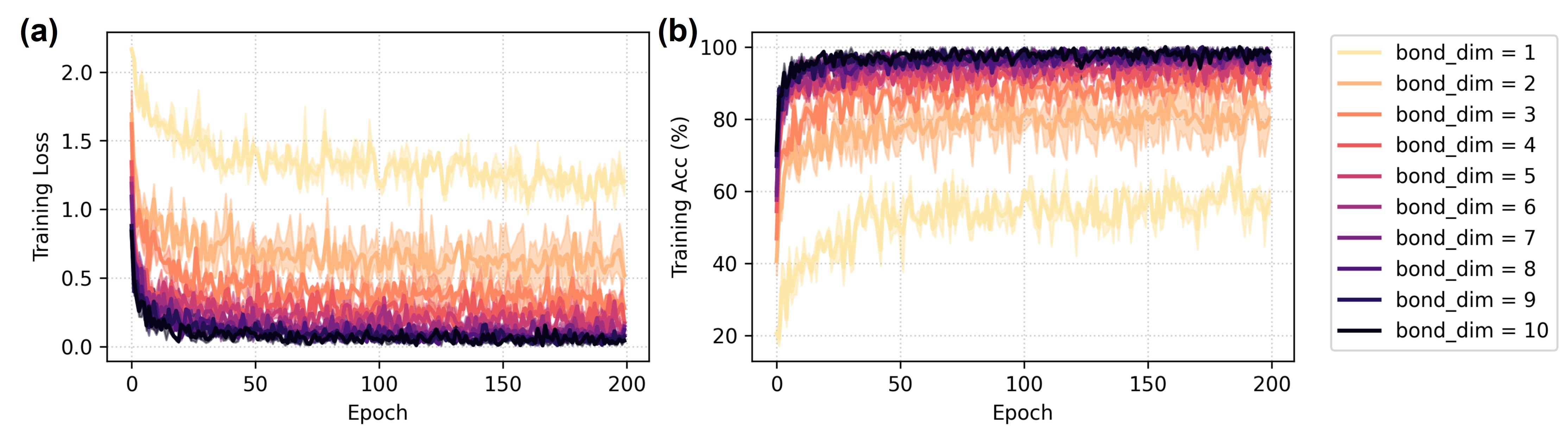}
\caption{
Training metrics of the photonic QT framework. (a) Training loss versus epochs for different bond dimensions. (b) Training accuracy versus epochs for different bond dimensions. Higher bond dimensions lead to lower loss and higher accuracy, underscoring the enhanced representational capacity of the photonic QT approach.
}
\label{fig:train_loss_acc}
\end{figure*}

\subsubsection{Weight Sharing}
Weight sharing is one of the more straightforward techniques for reducing the effective number of parameters in a neural network, whereby certain weights are reused across different parts of the model \cite{neill2020overviewneuralnetworkcompression}. While advanced strategies involve clustering pre-trained weights or incorporating specialized penalty terms, such approaches typically require a pre-trained model to identify the weights most amenable to sharing. In this work, we implement a simpler scheme that dispenses with any pre-training phase.

Consider an original weight matrix 
\[
W \in \mathbb{R}^{m \times n},
\]
where \(m\) denotes the number of input features and \(n\) the number of output features. We introduce a smaller collection of shared vectors
\[
\mathbf{v}_k \in \mathbb{R}^n, \quad k = 1, 2, \ldots, K,
\]
where \(K < m\). The idea is to replace each row of \(W\) with one of these shared vectors. Concretely, for each row index \(i \in \{1, 2, \ldots, m\}\), we define
\[
W_{i,j} \;=\; \mathbf{v}_{\sigma(i),j},
\]
where \(\sigma: \{1, 2, \ldots, m\} \to \{1, 2, \ldots, K\}\) is a mapping function (in our simplest implementation, we choose a regular assignment of rows to each shared vector in sequence), and \(j \in \{1, 2, \ldots, n\}\). By tuning the number \(K\) of distinct shared vectors, one may control the total parameter count:
\[
\text{\# of parameters} \;=\; K \times n.
\]
This form of weight sharing thus provides a direct and intuitive trade-off between model size and performance.

In Fig.~\ref{fig:acc_gen_error} (blue triangles), we plot the average testing accuracy (over three independent runs) for varying numbers of trainable parameters under the weight-sharing scheme. Although increasing the number of shared vectors generally improves accuracy, we observe that performance saturates below that of the photonic QT approach for comparable parameter counts. Nonetheless, weight sharing remains a competitively simple and efficient baseline for parameter reduction in purely classical contexts.

\subsubsection{Pruning}
Pruning is arguably one of the most extensively studied methods for reducing parameter counts in neural networks. Typically, pruning operates on a pre-trained network by removing weights deemed unimportant, thereby yielding a more compact representation that can then be retrained for further optimization. This process not only reduces storage demands but can also accelerate inference, especially when hardware supports structured pruning.

To ensure a fair comparison with the photonic QT framework, which does not rely on pre-training, we adopt a random pruning procedure from the outset. Specifically, for each trial, we remove a fraction \(\alpha\) of the parameters uniformly at random; thus, the resulting parameter count is \((1 - \alpha) \times (m \times n)\). We then train this pruned network from scratch. To account for the variability introduced by random selection, we repeat each experiment three times and report the average performance.

In Fig.~\ref{fig:acc_gen_error} (red circles), we show the testing accuracy under various levels of pruning. While moderate pruning achieves competitive performance, a high pruning ratio can degrade accuracy. Even so, pruning remains a strong baseline approach for model compression and can be considered alongside photonic QT, especially in resource-constrained, fully classical settings.

\subsection{Model Compression and Parameter Efficiency}
\label{sec:model_compression}

To evaluate our photonic QT framework under the Quandela challenge---a classification task on a partial MNIST dataset---we target a classical convolutional neural network (CNN) with a total of \(m = 6690\) parameters. In alignment with the photonic QT scheme described above, we employ two photonic QNNs of mode and qubit configurations \(M_1 = 9, N_1 = 4, M_2 = 8, N_2 = 4\). This setup features \(108 + 84 = 192\) QNN parameters, enabling the generation of \(C(M_1, N_1) \times C(M_2, N_2) = 8820\) parameters, among which \(6690\) are then selected to construct the weights of the target classical CNN. Additionally, the matrix product state (MPS) mapping model incorporates further parameters tied to its bond dimension, which we vary from 1 to 10.

\begin{table*}[!t]
    \centering
    \caption{Results of photonic QT with different bond dimension settings of MPS mapping model.}
    \vspace{10pt} 
    \small
    \begin{tabular}{ccccc}
        \toprule
          \textbf{Bond} & \textbf{\# of training} & \textbf{Training} & \textbf{Testing} & \textbf{Generalization} \\
          \textbf{dimension} & \textbf{parameters} & \textbf{accuracy (\%)} & \textbf{accuracy (\%)} & \textbf{error} \\
        \midrule
         1 & 223 & 58.256 $\pm$ 2.34 & 55.775 $\pm$ 3.27 & 0.0219 $\pm$ 0.007\\
         2 & 316 & 83.340 $\pm$ 2.77 &  81.375 $\pm$ 2.28 & 0.0462 $\pm$ 0.032\\
         3 & 471 & 88.693 $\pm$ 1.67 &  87.057 $\pm$ 2.66 & 0.0364 $\pm$ 0.016\\
         4 & 688 & 93.916 $\pm$ 0.45 &  93.292 $\pm$ 0.62 & 0.0679 $\pm$ 0.002\\
         5 & 967 & 95.450 $\pm$ 0.39 &  93.042 $\pm$ 0.77 & 0.0950 $\pm$ 0.010\\
         6 & 1308 & 96.953 $\pm$ 0.02 &  94.917 $\pm$ 0.60 & 0.1135 $\pm$ 0.013\\
         7 & 1711 & 97.773 $\pm$ 0.22 &  94.957 $\pm$ 0.82 & 0.1315 $\pm$ 0.031\\
         8 & 2176 & 97.866 $\pm$ 0.78 &  94.707 $\pm$ 0.47 & 0.1399 $\pm$ 0.007\\
         9 & 2703 & 98.373 $\pm$ 0.12 &  94.835 $\pm$ 0.48 & 0.1624 $\pm$ 0.021\\
         10 & 3292 & 98.990 $\pm$ 0.34 &  95.502 $\pm$ 0.84 & 0.2552 $\pm$ 0.053\\
        \bottomrule
    \end{tabular}
    \label{tab:diff_bd_results}
\end{table*}

\begin{figure*}[!t]
\centering
\includegraphics[width=\linewidth]{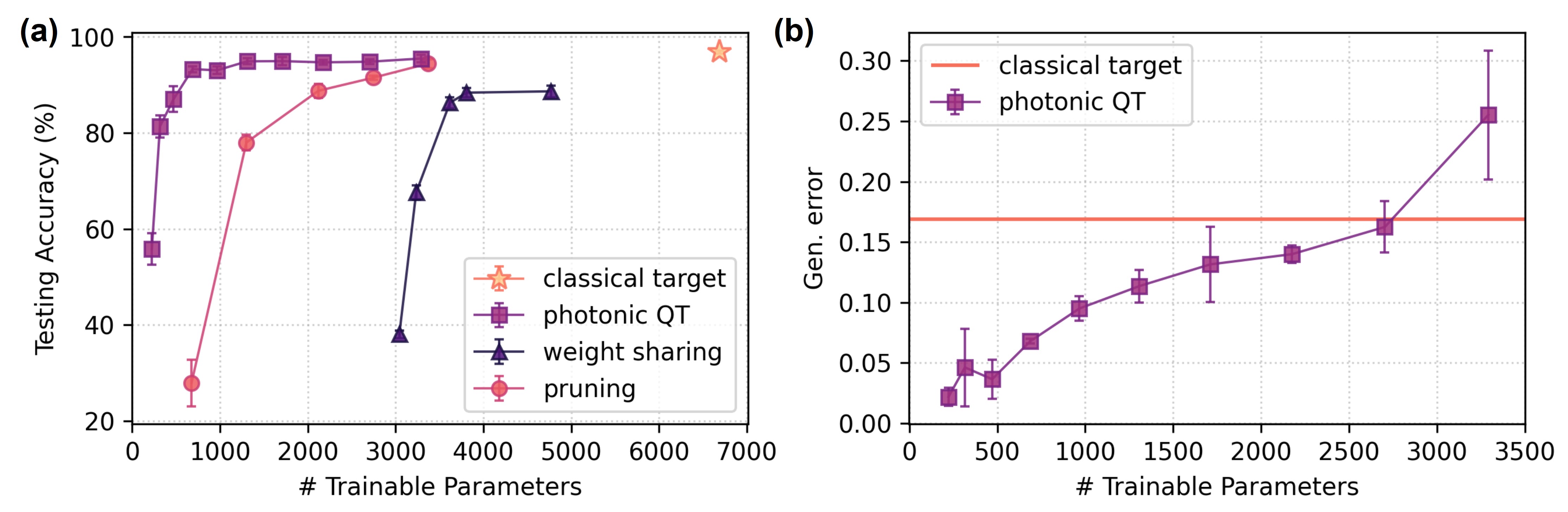}
\caption{
\textbf{Testing accuracy and generalization error across parameter-efficient training methods.} (\textbf{a}) Testing accuracy for photonic QT, classical weight sharing, pruning, and the original CNN. (\textbf{b}) Generalization error of photonic QT compared to the original CNN baseline. Photonic QT achieves competitive accuracy with significantly fewer parameters, albeit with an increased generalization error as model size grows.
}
\label{fig:acc_gen_error}
\end{figure*}

Fig.~\ref{fig:train_loss_acc} illustrates the training loss and accuracy for different MPS bond dimensions, tracked over 200 epochs. The left panel shows the progressive decrease in training loss for all tested bond dimensions. Notably, higher bond dimensions yield lower final loss values, indicating superior optimization and representational power. In parallel, the right panel highlights a corresponding rise in training accuracy; once again, higher bond dimensions result in improved final accuracy values.

Table~\ref{tab:original_cnn} summarizes the baseline performance of the original classical CNN. This model uses all 6690 trainable parameters and achieves a training accuracy of \(99.983 \pm 0.02\%\), a testing accuracy of \(96.890 \pm 0.31\%\), and a generalization error of \(0.1690 \pm 0.005\). Table~\ref{tab:diff_bd_results} reports the results for the photonic QT framework with bond dimensions ranging from 1 to 10. As the bond dimension increases, the number of trainable parameters grows from 223 to 3292, leading to marked improvements in both training and testing accuracies. For instance, with a bond dimension of 1, the training and testing accuracies are \(58.256 \pm 2.34\%\) and \(55.775 \pm 3.27\%\), respectively. By contrast, a bond dimension of 10 achieves \(98.990 \pm 0.34\%\) training accuracy and \(95.502 \pm 0.84\%\) testing accuracy.

While the higher bond dimensions confer substantial gains in accuracy, they also raise the generalization error. At a bond dimension of 1, the generalization error is \(0.0219 \pm 0.007\), but grows to \(0.2552 \pm 0.053\) at a bond dimension of 10. This highlights a critical trade-off in the photonic QT approach between the benefits of higher model capacity and the risks of overfitting.

Fig.~\ref{fig:acc_gen_error} compares testing accuracy and generalization error for the photonic QT framework against classical compression baselines (weight sharing and pruning) as well as the original CNN. The left panel displays the testing accuracy as a function of the number of trainable parameters. Although the classical target model attains the highest testing accuracy (\(\sim 96\%\)) with 6690 parameters, photonic QT approaches this accuracy with fewer parameters, thus demonstrating superior parameter efficiency. Weight sharing (blue triangles) shows rapid accuracy gains at low parameter counts but plateaus below photonic QT levels; pruning (red circles) continues to improve over a broader parameter range, eventually converging toward the photonic QT performance.

\begin{table*}[!t]
    \centering
    \caption{Parameter Efficiency Benchmarking: Comparative Analysis of Training Parameter Counts and Testing Accuracy Across Compression Methods on MNIST Classification}
    \vspace{10pt} 
    \small
    \begin{tabular}{cccc}
        \toprule
           \textbf{Method}  & \textbf{\# of training parameters} & \textbf{Testing accuracy (\%)} \\
        \midrule
          Original & 6690 & 96.890 $\pm$ 0.31 \\
          Weight sharing  & 4770 & 88.666 $\pm$ 1.207 \\
          Pruning & 3370 & 94.443 $\pm$ 0.923 \\
          Photonic QT (bond dimension $=10$)& 3292  & 95.502 $\pm$ 0.84 \\
          Photonic QT (bond dimension $=4$)& 688  & 93.292 $\pm$ 0.62 \\

        \bottomrule
    \end{tabular}
    \label{tab:mm}
\end{table*}

The left panel highlights the testing accuracy trends across different methods. The classical target model achieves the highest testing accuracy ($\sim 96\%$) but requires a significantly larger number of trainable parameters ($6690$), serving as a baseline for comparison. In contrast, the photonic QT framework demonstrates a clear improvement in testing accuracy as the number of trainable parameters increases. Remarkably, photonic QT achieves testing accuracy comparable to the classical target model while requiring far fewer parameters, showcasing its potential for parameter-efficient training.

The weight-sharing method, depicted as blue triangles, shows rapid gains in testing accuracy with increasing parameter counts. However, it plateaus at slightly lower accuracy levels compared to photonic QT, highlighting its limitations in fully capturing complex patterns. The pruning method, represented by red circles, starts with lower testing accuracy for smaller parameter counts but shows consistent improvement as the number of trainable parameters increases, eventually approaching the performance of photonic QT for larger models.

The right panel of Fig.~\ref{fig:acc_gen_error} focuses on generalization error. While the original CNN achieves a generalization error of \(0.169\), photonic QT exhibits a growing generalization error as its model size increases, eventually exceeding that of the classical baseline. Consequently, although photonic QT offers compelling parameter-efficiency benefits, particularly at lower bond dimensions, careful tuning is required to curb overfitting and maintain robust generalization.

Table~\ref{tab:mm} further underscores the unique advantages of photonic QT over classical compression methods in terms of parameter efficiency and testing accuracy. Nevertheless, weight sharing and pruning remain viable alternatives for reducing parameter counts when quantum resources are unavailable. Overall, these findings illustrate the promise of the photonic QT framework in delivering high testing accuracy with comparatively few trainable parameters, while also highlighting the ongoing need to manage overfitting in quantum-enhanced machine learning.

\begin{figure*}[!t]
\centering
\captionsetup[subfloat]{labelfont=large, textfont=normalsize} % Adjust label size here
\subfloat[Brightness]{\includegraphics[width=0.248\linewidth]{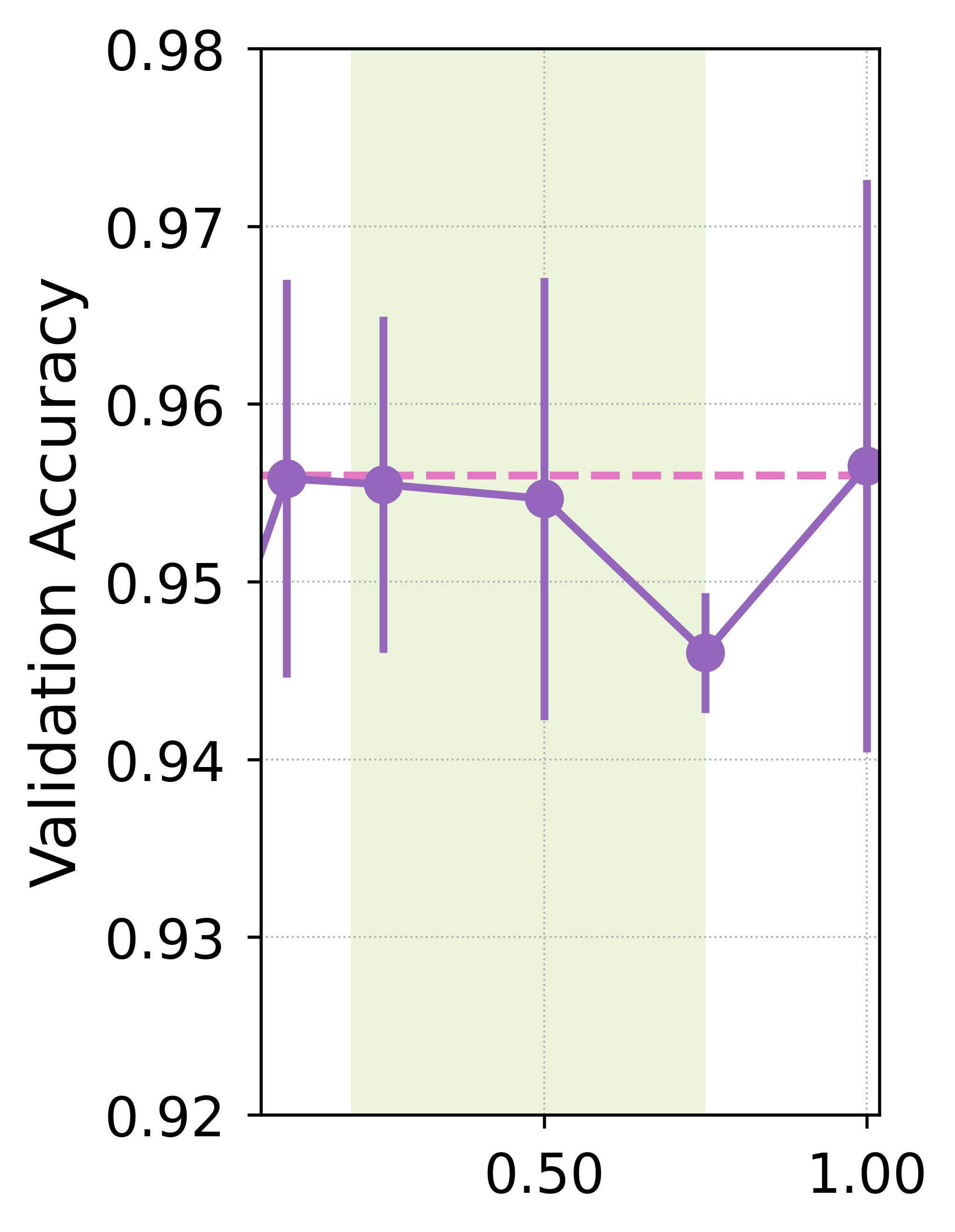}} \hfill
\subfloat[Indistinguishability]{\includegraphics[width=0.24\linewidth]{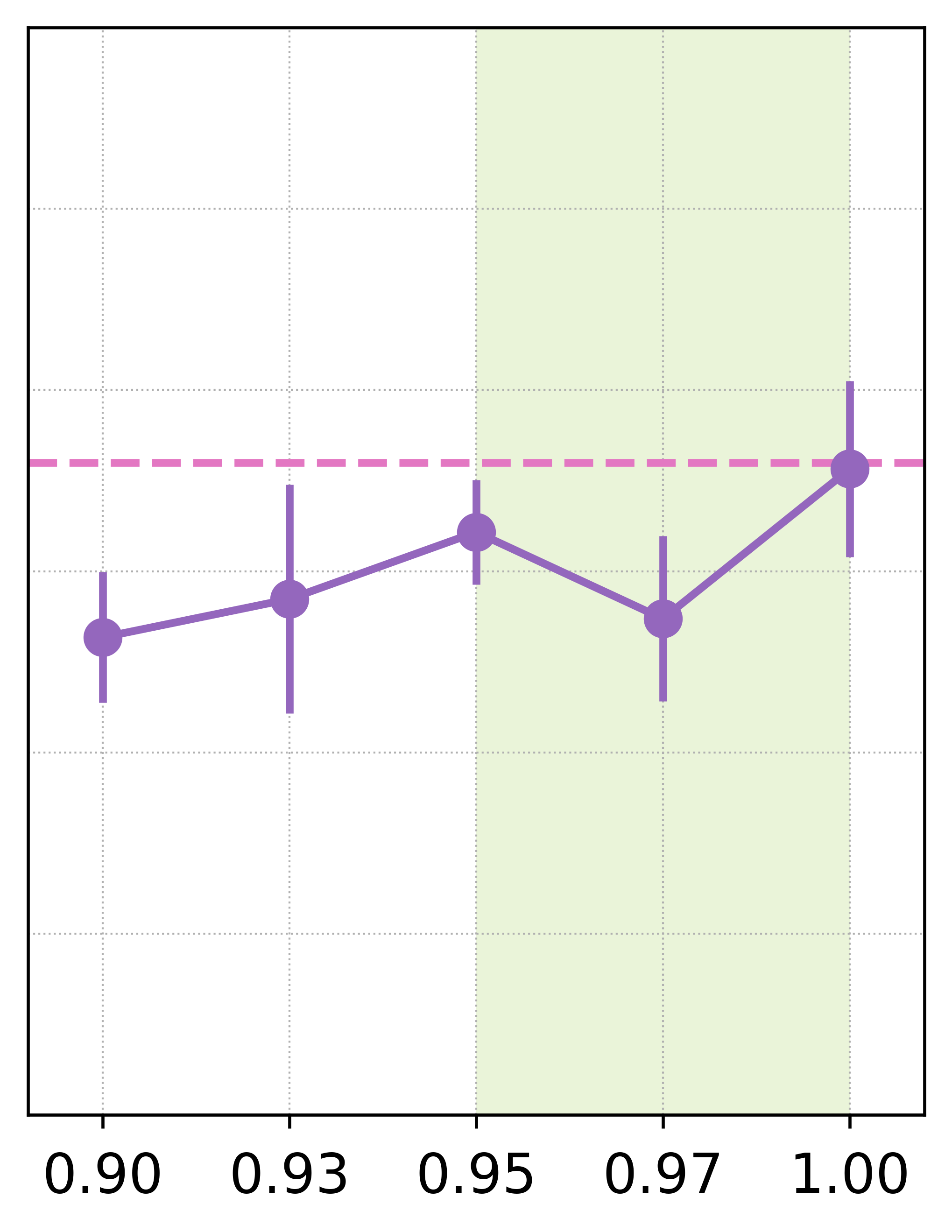}} \hfill
\subfloat[Second-order Correlation]{\includegraphics[width=0.24\linewidth]{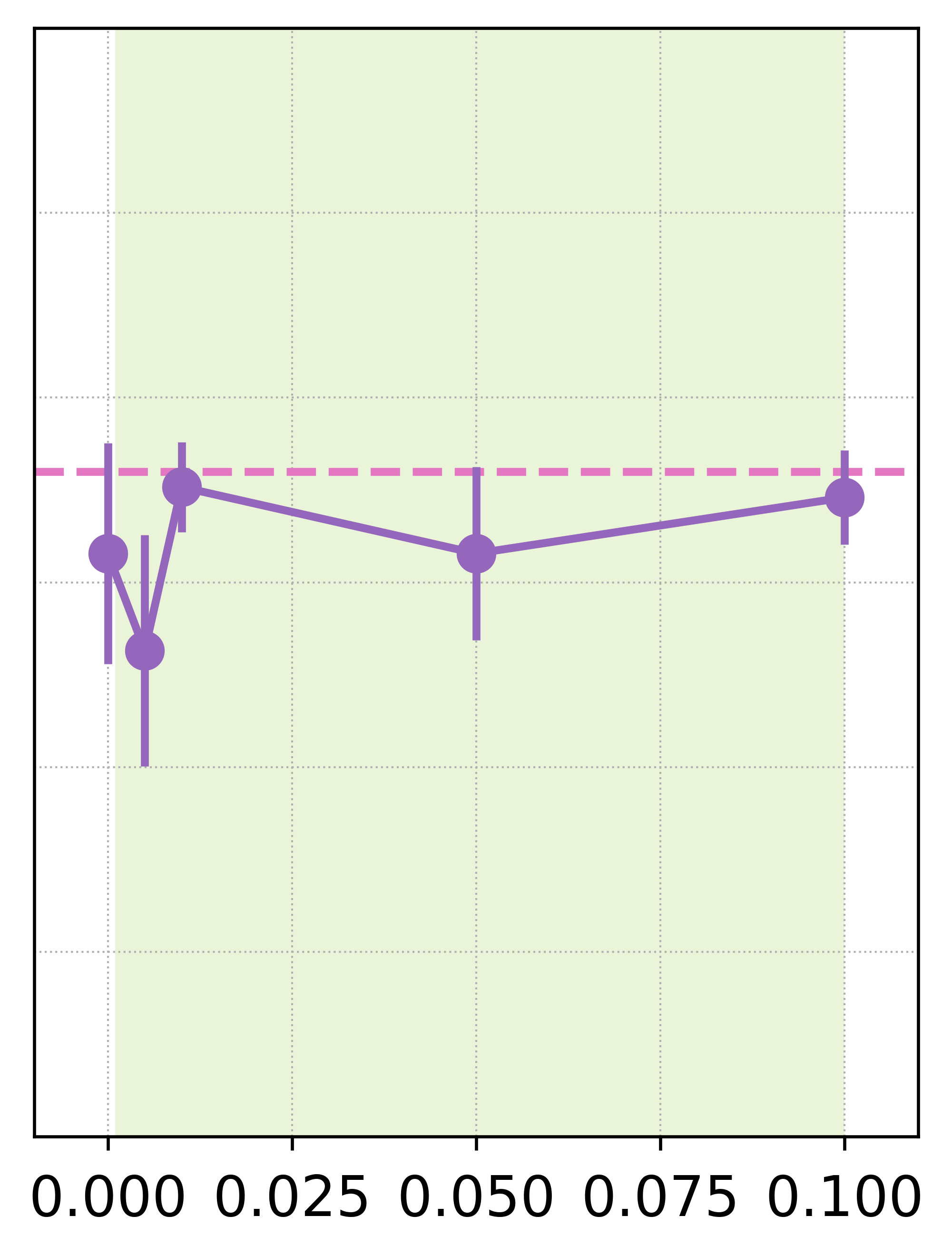}} \hfill
\subfloat[Transmittance]{\includegraphics[width=0.24\linewidth]{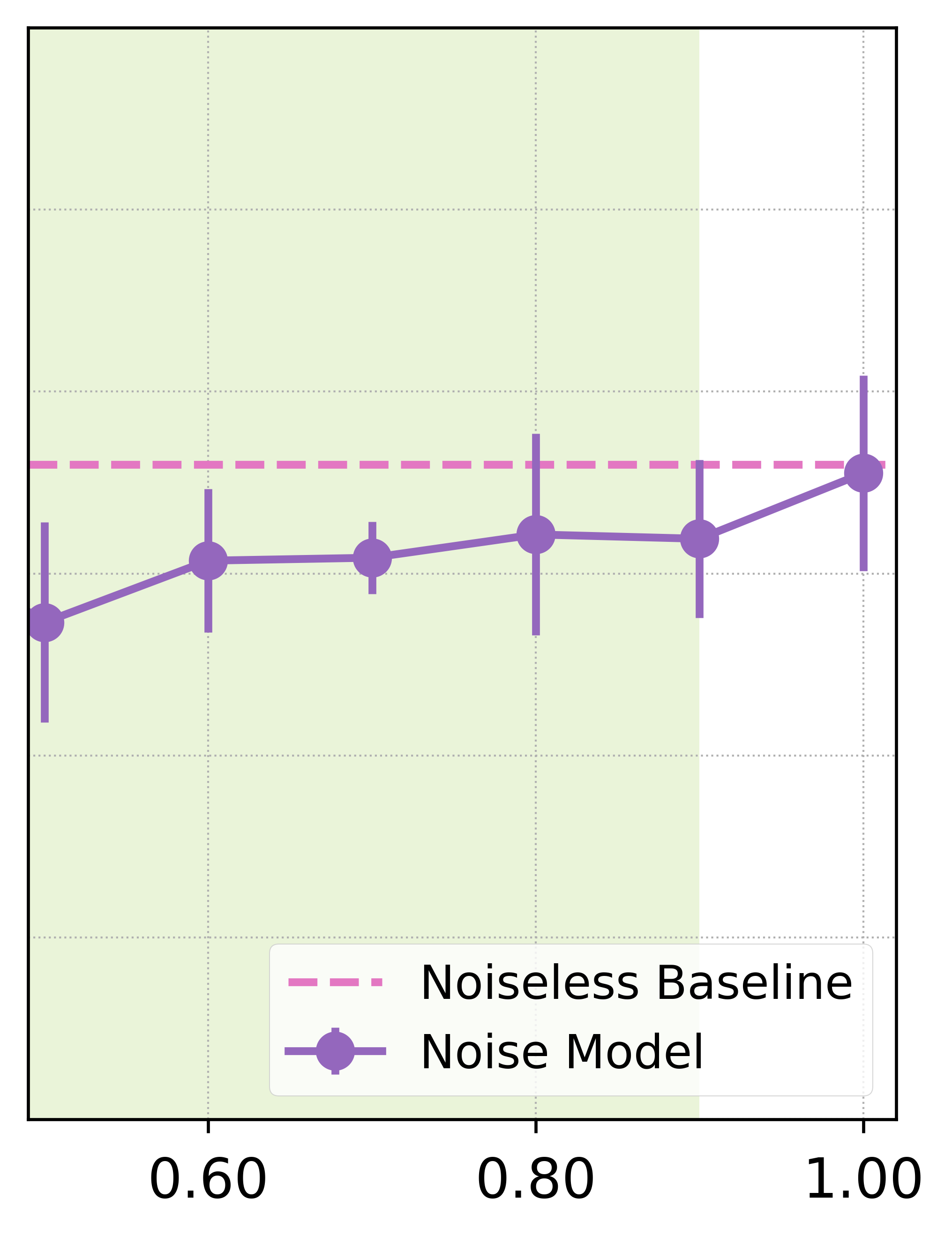}}

\caption{Validation accuracy of the hybrid photonic–classical model under four elementary noise channels. Each curve is the mean of three independent training runs and is compared against an ideal (noiseless) baseline obtained with the same optimizer hyperparameters. Subfigures correspond to: (a)~brightness $\beta$, (b)~indistinguishability $I$, (c)~second-order correlation $g^{(2)}(0)$, and (d)~transmittance $T$. The green shaded region indicates the range of realistic noise parameters achievable in the current state of the art (see Table~\ref{tab:noise_params}).}
\label{fig:noise-sweep}
\end{figure*}

\begin{table*}[!t]
  \centering
  \caption{Key hardware parameters for state-of-the-art single-photon sources
           and integrated photonic circuits.}
  \label{tab:noise_params}
  \begin{tabular}{@{}lccc@{}}
    \toprule
    Symbol            & Figure of merit                 & Physical meaning
                      & Typical value \\ \midrule
    $\beta$           & Brightness                      & Single-photon emission
                                                          probability per clock
                                                          cycle
                      & 0.2–0.75\cite{maring2024versatile} \\
    $I$               & Indistinguishability            & Wave-packet overlap
                                                          (HOM visibility)
                      & $>0.95$\cite{ollivier2021hong} \\
    $g^{(2)}(0)$      & Second-order correlation        & Multiphoton emission
                                                          probability
                      & $10^{-1}$–$10^{-3}$\cite{chen2024heralded} \\
    $T$               & Transmittance                   & Probability that a
                                                          photon survives
                                                          propagation, coupling
                                                          and detection
                      & 0.4-0.9\cite{chanana2022ultra,yang2025programmable} \\
    \bottomrule
  \end{tabular}
\end{table*}

\subsection{Noise Resilience of Photonic Quantum-Train}
Photonic qubits interact predominantly via linear–optical interference, so computational accuracy is limited less by coherence time and more by \emph{source quality} and \emph{optical loss}. We therefore adopt an experimentally realistic model that captures the four hardware figures of merit that dominate current photonic platforms: brightness~($\beta$), indistinguishability~($I$), second-order correlation~($g^{(2)}(0)$) and system transmittance~($T$). All remaining parameters are kept at their ideal values so that the logical error budget can be attributed unambiguously to these four channels.

\subsubsection{Brightness}
\label{subsec:brightness}

In photonic quantum information, the brightness of a pulsed single-photon source refers to the probability that a single excitation pulse generates exactly one photon in the target mode, accounting for all collection, coupling, and filtering losses. This metric reflects the deterministic quality of the emitter and sets a fundamental limit on the scalability of measurement-based photonic quantum computing architectures\cite{ding2025high}.

In distributed quantum architectures requiring $N$-photon coincidence events, the expected rate of successful gate operations scales as~$\beta^N$ due to the independence of single-photon sources. Insufficient brightness induces \emph{vacuum-dominated noise} where gate failures occur probabilistically when photon number statistics deviate from the required Fock-state inputs.

For our neural-network architecture, the heralding mechanism provides crucial error discrimination: vacuum events in any optical channel result in protocol abortion rather than faulty computation. This transforms the brightness constraint into an exponential resource overhead $\mathcal{O}(\beta^{-N})$ for successful circuit execution, as opposed to the polynomial overheads associated with logical error correction. The separation between temporal resource costs and state fidelity preservation represents a key advantage of our heralded approach over conventional non-deterministic photonic architectures.

\subsubsection{Indistinguishability}
\label{subsec:indistinguishability}

The indistinguishability \( I \equiv |\langle \psi_1 | \psi_2 \rangle|^2 \) quantifies the wavefunction overlap between single-photon states from independent sources, with $I=1$ characterizing perfect bosonic interference capability.

Two-photon interference processes -- the foundation of linear optical quantum computing -- achieve ideal Hong-Ou-Mandel (HOM) visibility $V_0=1$ when $I=1$. Practical implementations exhibit reduced visibility
\begin{equation}
  V = I\left(1 - g^{(2)}(0)\right),
  \label{eq:visibility}
\end{equation}
where $g^{(2)}(0)$ denotes the second-order correlation function at zero delay. For single-photon Fock states ($g^{(2)}(0)=0$), the visibility directly measures $I$. This degradation propagates to gate fidelity as
\begin{equation}
  F_{\text{gate}} \approx \frac{1}{2}(1 + V),
  \label{eq:fidelity}
\end{equation}
establishing $I$ as a critical performance parameter.

In distributed architectures where photons originate from independent emitter modules, spectral diffusion from locally fluctuating pump lasers and temperatures induces time-dependent phase mismatches $\phi(t)$. The resulting averaged indistinguishability
\begin{equation}
  \bar{I} = \left|\int dt \braket{\psi_1(t)}{\psi_2(t)}\right|^2
\end{equation}
becomes the dominant non-loss error channel, exceeding other decoherence mechanisms. Our noise model incorporates this effect through a depolarizing channel $\mathcal{E}(\rho) = (1-p)\rho + p\frac{\mathbb{I}}{d}$ acting on each attempted interference, with error probability $p \propto 1-\bar{I}$ derived from measured spectral wandering statistics.

\subsubsection{Multiphoton contamination}
\label{subsec:g2}

The second-order correlation function $g^{(2)}(0) \equiv \frac{\langle \hat{n}(\hat{n}-1) \rangle}{\langle \hat{n} \rangle^2}$ quantifies multiphoton emission probability, where $\hat{n}$ is the photon number operator. Ideal single-photon sources exhibit $g^{(2)}(0)=0$.

Multiphoton events introduce \emph{Fock-state leakage} in boson-sampling architectures, contaminating the computational basis with $\ket{n\geq2}$ states that disrupt quantum interference patterns. The resulting logical error rate per $N_{\text{modes}}$-mode gate operation follows
\begin{equation}
  \varepsilon_{\text{multi}} \approx \frac{1}{2} N_{\text{modes}} g^{(2)}(0),
  \label{eq:multiphoton_error}
\end{equation}
where the $\frac{1}{2}$ factor arises from partial post-selection against photon-number discrepancies. This first-order approximation assumes (i) Poissonian statistics for multiphoton emissions and (ii) spatial mode independence.

% The quantity $g^{(2)}(0)$ measures the probability that a source emits two
% photons instead of one.
% Such multiphoton events introduce
% \emph{state leakage} into boson-sampling-style circuits.
% To first order, the logical error incurred per gate scales as
% %
% \begin{equation}
%   \varepsilon_{\text{multi}}
%   \simeq \tfrac12\,N_{\text{modes}}\,g^{(2)}(0),
% \end{equation}
% %
% where the factor~$\tfrac12$ accounts for post-selection that discards events
% with obvious photon-number mismatch.
% With $g^{(2)}(0)\lesssim0.03$ reported for modern quantum-dot sources,
% $\varepsilon_{\text{multi}}$ remains below the percent level for the circuit
% depths explored here.

\subsubsection{Transmittance}
\label{subsec:transmittance}
We treat the end-to-end optical transmittance as a single multiplicative parameter \(T\in[0,1]\) that captures fibre–chip coupling, on-chip propagation, routing, filtering, and detector quantum efficiency.  Recent progress in silicon nitride and hydex platforms has pushed waveguide loss below \(1\;\mathrm{dB/m}\) \cite{blumenthal2018silicon}, while cryogenic edge-coupling modules now exceed \(T_{\mathrm{fc}}\simeq0.90\) \cite{alexander2024manufacturable}.  Integrated superconducting nanowire detectors add \(<0.2\;\mathrm{dB}\) of insertion loss at \(>\!90\%\) system detection efficiency \cite{esmaeil2021superconducting}.  Combining these state-of-the-art figures yields a realistic device-level budget of \(T\approx0.8\) for a 10-cm photonic path, which sets the upper bound of the transmittance sweep used in our noise model; lower values emulate legacy platforms with higher coupling or propagation loss.

%Fidelity of operation 

\subsubsection{Noise Analysis}
Fig.\ref{fig:noise-sweep} establishes a hierarchy of physical error channels that aligns closely with the theoretical framework of Secs.\,\ref{subsec:brightness}–\ref{subsec:transmittance}.  
For brightness (\(\beta\)) the validation curve is essentially level: it starts at \(\approx0.955\) for \(\beta = 0\), stays within \(<1\text{\,ppt}\) of that value up to \(\beta \approx 0.50\), shows a transient dip to \(\approx0.945\) at \(\beta \approx 0.75\), and recovers to \(\approx0.955\) at \(\beta = 1\).  
The shallow excursion indicates that heralding inefficiencies and multi-pair contamination exert competing influences whose net effect remains below the \(1\%\) level across the full pump-power range. Varying the indistinguishability within \(0.90\le I\le1.00\) leaves the curve essentially flat (a dip of \(\lesssim0.01\) at \(I\approx0.93\) lies within optimiser variance), confirming that residual spectral-temporal mismatch acts only as a weak depolarising perturbation compared with brightness-induced Fock-state leakage.    The $g^{(2)}(0)$ sweep shows the complementary trade‑off: accuracy is lowest at strictly vanishing multi‑photon probability, climbs monotonically to $g^{(2)}(0)=0.10$, and flattens thereafter; a tiny impurity acts as label‑smoothing–like regularisation that enriches the training distribution, but once $g^{(2)}(0)\gtrsim0.05$ the deleterious two‑photon component dominates.  Optical loss exhibits the mildest influence: accuracy increases monotonically—by \(\sim7\times10^{-3}\)—as the transmittance is raised from \(T=0.6\) to \(T=1.0\); thus attenuation removes only a small fraction of useful single-photon events without materially suppressing residual multi-photon noise, and unit throughput essentially reproduces the noiseless benchmark.  Across all sweeps the worst‑case degradation is confined to less than three percentage points, identifying excess multi‑photon emission at high brightness as the principal residual error source and demonstrating that the hybrid photonic–classical architecture maintains high‑fidelity operation under first‑order imperfections realistic for current hardware.

\subsection{Ablation Study}
While our results demonstrate the superior parameter efficiency and noise-resilience of photonic QT, it is essential to investigate the origin of this enhanced performance. Specifically, we must discern whether the improvement fundamentally arises from the quantum architecture or can be attributed solely to classical components.

\begin{figure}[!b]
\centering
\includegraphics[width=\linewidth]{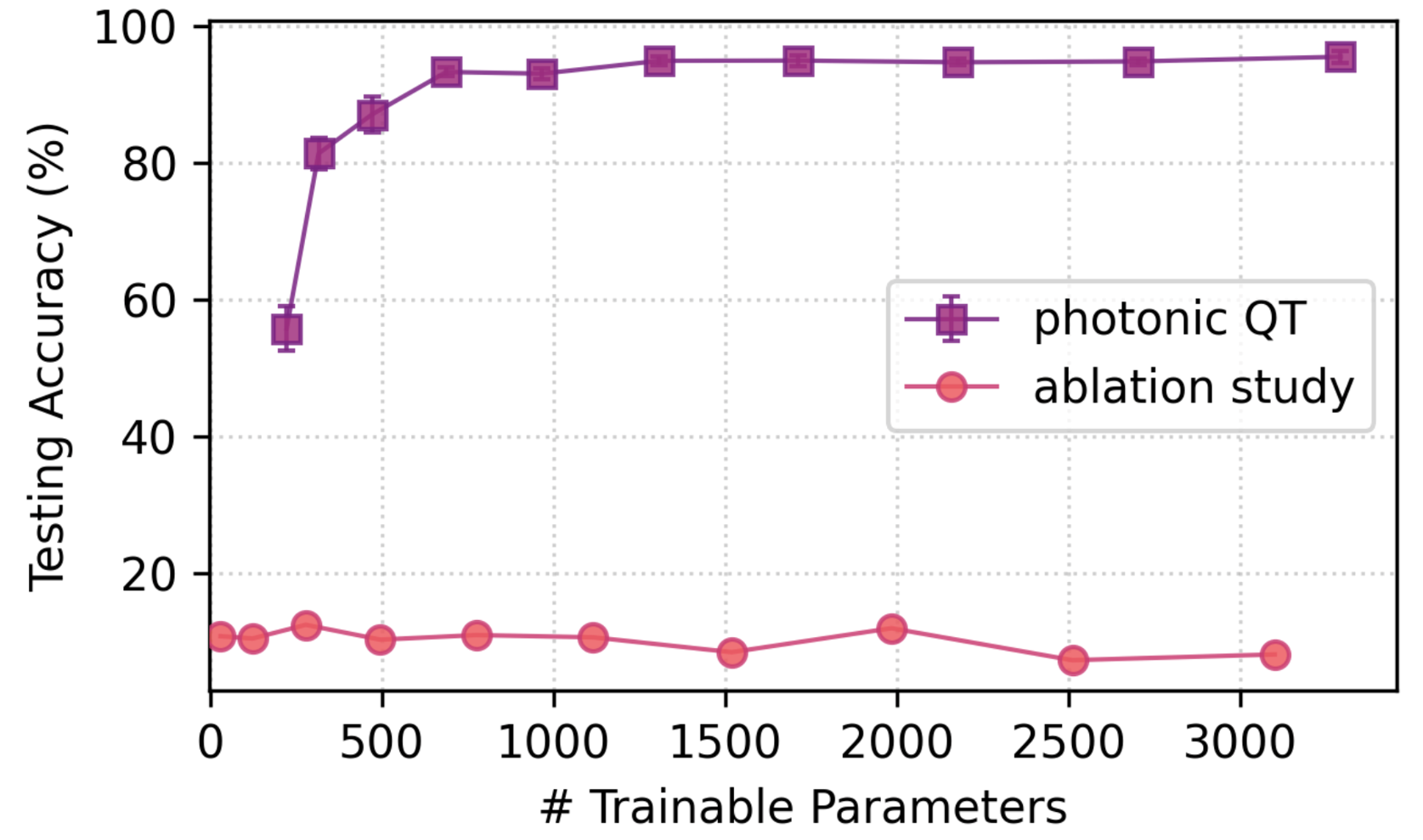}
\caption{Ablation analysis of photonic QT architecture. Replacement of quantum components with random noise inputs (orange) reduces MNIST classification accuracy to chance level, compared to intact quantum-classical hybrid system (blue). Shaded regions represent 95\% confidence intervals over 10 trials.}
\label{fig:ablation_study}
\end{figure}

Analysis of our parameter tables reveals that the majority of trainable parameters reside in the MPS mapping module. The observed correlation between increased MPS bond dimension ($\chi$) and improved performance raises a critical question: \textit{Does the MPS module possess sufficient expressive power to render the quantum components redundant?}

To test this hypothesis, we perform an ablation study where we replace all quantum components (the photonic quantum neural network and associated preprocessing stages, as shown left of the MPS in Fig.~\ref{fig:scheme}) with random noise vectors. In this modified architecture, the MPS effectively operates as a classical generator transforming stochastic inputs into weight matrices -- analogous to a conventional generative model. This configuration isolates the contribution of the classical MPS component.

The results presented in Fig.~\ref{fig:ablation_study} demonstrate catastrophic performance degradation when removing quantum components, with prediction accuracy collapsing to random guessing levels ($10.0\% \pm 0.5\%$ for 10-class MNIST classification). Crucially, this performance floor persists across all tested bond dimensions ($\chi = 2$ to $16$), establishing that the MPS alone cannot extract meaningful features without quantum-processed inputs. This definitive ablation study confirms the quantum component is indispensable for photonic QT's functionality.

\section{Conclusion and Future Work}
\label{sec:cfw}
This work establishes photonic QT as a scalable paradigm for quantum-enhanced neural network compression, bridging photonic quantum computing's theoretical advantages with practical classical deployment. By leveraging universal linear-optical interferometers ($U(M) \in \mathbb{C}^{M\times M}$) and matrix product state (MPS) mapping, our framework achieves $10\times$ parameter compression ($\chi=4$) with only $3.50\%$ relative accuracy loss on MNIST classification ($93.29\% \pm 0.62\%$ vs classical $96.89\% \pm 0.31\%$). The architecture's distributed design—partitioning $N$-photon $M$-mode QNNs across photonic processors—reduces per-node qubit requirements to $\mathcal{O}(NM)$ while maintaining room-temperature operation. Crucially, ablation studies confirm quantum necessity: replacing photonic components with random inputs collapses accuracy to $10.0\% \pm 0.5\%$, proving quantum state evolution generates irreplaceable features beyond classical MPS capabilities. The demonstrated parameter efficiency ($3,292$ vs $6,690$ parameters) and superior performance over classical compression methods (6-12\% accuracy gains) position photonic QT as a viable solution for quantum-machine learning interoperability in resource-constrained environments. Comprehensive noise analysis under realistic photonic imperfections demonstrates the framework’s noise resilience, with worst-case accuracy degradation confined to \(<3\%\) across all noise channels.

Future research will focus on three critical frontiers: (1) Optimizing the bond dimension-generalization trade-off through regularized MPS architectures with entanglement entropy constraints and enhanced quantum-classical co-design of parameterized quantum circuits; (2) Scaling to large artificial intelligence models by investigating photonic QT's applicability to transformer architectures and few-shot fine-tuning of language models under $\mathcal{O}(\log N)$ parameter growth regimes\cite{liu2024quantumiclr}; (3) Developing noise resilience protocols for distributed photonic computing, including photon-loss tolerant encodings and Bayesian inference techniques for measurement-based error correction across multi-node configurations. Realizing these advancements will require co-developing photonic hardware with dynamic reconfigurability and HPC-integrated control systems, ultimately enabling fault-tolerant quantum neural networks across computer vision, natural language processing, and scientific machine learning domains.

\section*{Code Availability}
All scripts used to reproduce the framework of this research and noise sweeps are available at \href{https://github.com/Louisanity/PhotonicQuantumTrain}{\textcolor{blue}{https://github.com/Louisanity/PhotonicQuantumTrain}}. The simulation backend for photonic quantum computing relies on \textsc{Perceval}, an open‑source library developed by Quandela and accessible at \href{https://github.com/Quandela/Perceval}{\textcolor{blue}{https://github.com/Quandela/Perceval}}.

\section*{Acknowledgment}
The author would like to thank Jean Senellart (Quandela) for the helpful discussions. This work was supported by the Engineering and Physical Sciences Research Council (EPSRC) under grant number EP/W032643/1 and Imperial QuEST Seed Fund.

\bibliographystyle{ieeetr}
\bibliography{references,ref_YC, bib/peft, bib/qml, bib/qt}

\end{document}